\documentclass[12pt]{iopart}

\usepackage{graphicx}
\usepackage{amsfonts}

\begin{document}

\title{Enhancement of cooperation by giving high-degree neighbors more help}

\author{Han-Xin Yang$^{1,2}$ and Zhi-Xi Wu$^{3}$}
\address{$^{1}$Department of Physics, Fuzhou University, Fuzhou
350116, China}
\address{$^{2}$Center for Discrete Mathematics, Fuzhou University,
Fujian, 350116, China}
\address{$^{3}$Institute of Computational Physics and Complex Systems,
Lanzhou University, Lanzhou, Gansu 730000, People's Republic of
China}

\ead{$^{1}$hxyang01@gmail.com}\ead{$^{3}$eric0724@gmail.com}
\begin{abstract}
In this paper, we study the effect of preferential assistance on
cooperation in the donation game. Cooperators provide benefits to
their neighbors at some costs. Defectors pay no cost and do not
distribute any benefits. The total contribution of a cooperator is
fixed and he/she distributes his/her contribution unevenly to
his/her neighbors. Each individual is assigned a weight that is the
power of its degree, where the exponent $\alpha$ is an adjustable
parameter. The amount that cooperator $i$ contributes to a neighbor
$j$ is proportional to $j$'s weight. Interestingly, we find that
there exists an optimal value of $\alpha$ (which is positive),
leading to the highest cooperation level. This phenomenon indicates
that, to enhance cooperation, individuals could give high-degree
neighbors more help, but only to a certain extent.

\end{abstract}


 \maketitle
\tableofcontents
\section{Introduction} \label{sec:intro}

Evolutionary game theory has been frequently employed as the
theoretical framework to explain the emergence of cooperation among
selfish individuals~\cite{game1}. One of the most used game model is
the prisoner's dilemma game (PDG)~\cite{pdg}. In PDG played by two
individuals, each one simultaneously decides whether to cooperate or
to defect. They both receive $R$ upon mutual cooperation and $P$
upon mutual defection. If one cooperates but the other defects, the
defector gets the payoff $T$, while the cooperator gains the payoff
$S$. The payoff rank for PDG is $T > R > P > S$. With the rapid
development of network science, studies of PDG and other game models
are implemented on various
networks~\cite{spatial1,spatial2,spatial3,spatial4,spatial5,spatial6},
including regular
lattices~\cite{lattice0,lattice1,lattice2,lattice3,lattice4,lattice5,lattice6,lattice7},
random graphs~\cite{random1,random2}, small-world
networks~\cite{small1}, scale-free
networks~\cite{santos1,santos2,scale0,scale1,scale2,scale3},
multiplex networks~\cite{multi1,multi2} and so on. For a given
network, nodes represent individuals and links reflect social
relationships. Individuals play the PDG with their direct neighbors.

An important special case of the PDG is the so-called donation game,
where a cooperator provides a benefit $b$ to the other player at
his/her cost $c$, with $0 < c < b$. A defector pays no cost and does
not distribute any benefits. Thus, the payoff parameters in the
donation game are $T = b$, $R = b - c$, $P = 0$, and $S = -c$.
Ohtsuki $et$. $al$. discovered that natural selection favors
cooperation if the benefit-to-cost $b/c$ exceeds the average number
of neighbors~\cite{Ohtsuki}. Allen $et$. $al$. provided a solution
for weak selection that applies to any network and found that
cooperation flourishes most in societies which are based on strong
pairwise ties~\cite{Allen}. Wu $et$. $al$. investigated impact of
heterogeneous activity and community structure on the donation
game~\cite{Wu}. Hilbe $et$. $al$. showed that in large, well-mixed
populations, extortion strategies can play an important role, but
only as catalyzers for cooperation and not as a long-term
outcome~\cite{Hilbe}. Szolnoki and Perc found that extortion is
evolutionarily stable in structured populations if the strategy
updating is governed by a myopic best response rule~\cite{Szolnoki}.
Xu $et$. $al$. discovered that extortion strategies can act as
catalysts to promote the emergence of cooperation in structured
populations via different mechanisms~\cite{Xu}.

In previous studies of the spatial donation game, a cooperator
treats all neighbors equally and contributes to each neighbor with
the same cost. However, in real life, an individual usually has a
preference for somebody and provides more benefit to him/her. An
example is that altruistic act happens more frequently among genetic
relatives~\cite{Hamilton}. In this paper, we propose a heterogeneous
donation game in which a cooperator helps one of his/her neighbors
with the cost proportional to the neighbor's weight. We assign each
individual $i$ a weight $k_{i}^{\alpha}$, where $k_{i}$ is $i$'s
degree and $\alpha$ is an adjustable parameter. We have found that,
the cooperation level can be maximized at an optimal value of
$\alpha$.

\section{Model} \label{sec:methods}

A cooperator $i$ provides a benefit $rc_{ij}$ to a neighbor $j$ at a
cost $c_{ij}$, where $r$ is the benefit-to-cost ratio. For
simplicity, we assume that $r$ is the same for all pair
interactions. The total cost of cooperator $i$ is
\begin{equation} \label{eq1}
c_{i}=\sum _{j\epsilon \Omega_{i}}c_{ij},
\end{equation}
where the sum runs over all the direct neighbors of $i$ (this set is
indicated by $\Omega_{i}$). A defector pays no cost and does not
distribute any benefits.

Each individual $i$ is assigned a weight $k_{i}^{\alpha}$, where
$k_{i}$ is $i$'s degree and $\alpha$ is an adjustable parameter. For
a fixed $c_{i}$, cooperator $i$ helps one of its neighbors $j$ with
a cost $c_{ij}$ proportional to $j$'s weight
\begin{equation}\label{eq2}
c_{ij}=c_{i}\frac{k_{j}^{\alpha}}{\sum_{l\epsilon \Omega_{i}}
k_{l}^{\alpha}}.
\end{equation}
For $\alpha>0(<0)$, cooperators give high-degree (low-degree)
neighbors more help. In the case of $\alpha=0$, a cooperator
provides the same benefit to each neighbor.

The payoff of individual $i$ is given by
\begin{equation}\label{eq3}
M_{i}=-c_{i}s_{i}+r\sum_{j\epsilon \Omega_{i}}c_{ji}s_{j},
\end{equation}
where $s_{i}=1$ if $i$ is a cooperator and $s_{i}=0$ if $i$ is a
defector. Note that the donation is not symmetric, i.e., $c_{ij}
\neq c_{ji}$.

After each time step, all individuals synchronously update their
strategies as follows. Each individual $i$ randomly chooses a
neighbor $j$ and adopts $j$'s strategy with the
probability~\cite{fermi}

\begin{equation}\label{eq4}
W(s_{i}\leftarrow s_{j})=\frac{1}{1+\exp[(M_i-M_j)/\beta]},
\end{equation}
where the parameter $\beta$ ($>0$) characterizes noise to permit
irrational choices. As the noise $\beta$ decreases, the individuals
become more rational, i.e., they follow the strategies of neighbors
who have obtained higher payoffs with greater probabilities.

\section{Results and analysis} \label{sec:main results}

We carry out our model in Barab\'{a}si-Albert (BA) scale-free
network~\cite{ba} with size $N=5000$. Without loss of generality, we
assume that each cooperator pays the same total cost ($c_{i}=1$ for
any cooperator $i$). Initially, the two strategies, cooperation and
defection, are randomly distributed among the individuals with the
equal probability 1/2. The equilibrium fraction of cooperators
$\rho_{c}$ is obtained by averaging over the last $10^{4}$ Monte
Carlo time steps from a total of $10^{5}$ steps. Each data point
results from 20 different network realizations with 10 runs for each
realization.

\begin{figure}
\begin{center}
\includegraphics[width=100mm]{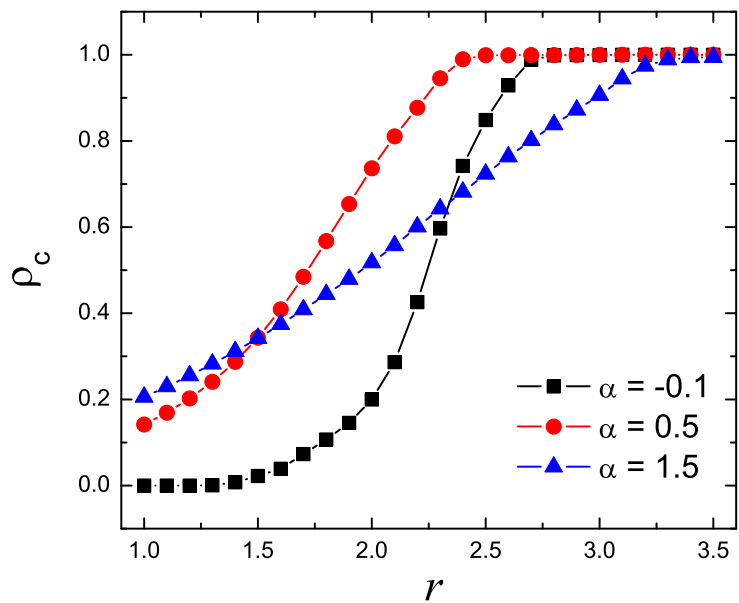}
\caption{(Color online) The fraction of cooperators $\rho_{c}$ as a
function of the benefit-to-cost ratio $r$ for different values of
$\alpha$. The average degree $\langle k \rangle$ = 10 and the noise
$\beta$ = 0.5. For each $\alpha$, $\rho_{c}$ increases to 1 as $r$
increases. }\label{fig1}
\end{center}
\end{figure}

Figure~\ref{fig1} shows the fraction of cooperators $\rho_{c}$ as a
function of the benefit-to-cost ratio $r$ for different values of
$\alpha$. One can see that for each value of $\alpha$, $\rho_{c}$
increases to 1 as $r$ increases. For a small value of $\alpha$
(e.g., $\alpha=-0.1$), cooperators die out when $r$ is close to 1.
However, for a large value of $\alpha$ (e.g., $\alpha$ = 0.5 or
$\alpha$ = 1.5), cooperators can still survive, even if $r=1$.

\begin{figure}
\begin{center}
\includegraphics[width=150mm]{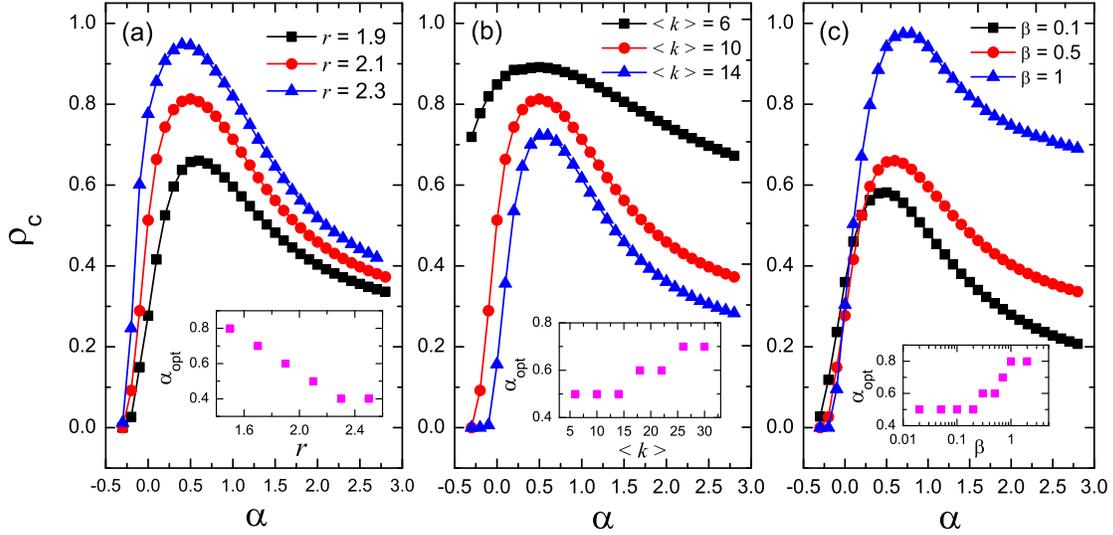}
\caption{(Color online) The fraction of cooperators $\rho_{c}$ as a
function of $\alpha$ for different values of (a) the benefit-to-cost
ratio $r$, (b) the average degree $\langle k \rangle$ and (c) the
noise $\beta$, respectively. For (a), $\langle k \rangle=10$ and
$\beta=0.5$. For (b), $r=2.1$ and $\beta=0.5$. For (c), $r=1.9$ and
$\langle k \rangle=10$. For fixed values of other parameters, there
exists an optimal value of $\alpha$ (denoted as $\alpha_{opt}$),
leading to the maximal $\rho_{c}$. The insets show the dependence of
$\alpha_{opt}$ on $r$, $\langle k \rangle$ and $\beta$ respectively.
} \label{fig2}
\end{center}
\end{figure}

Figure~\ref{fig2} shows the dependence of $\rho_{c}$ on $\alpha$.
One can see that for fixed values of other parameters, there exists
an optimal value of $\alpha$ (denoted as $\alpha_{opt}$), leading to
the maximal $\rho_{c}$. The value of $\alpha_{opt}$ is not fixed.
From the insets of Fig.~\ref{fig2}, one can see that $\alpha_{opt}$
decreases as $r$ increases, but increases as the average degree
$\langle k \rangle$ or the noise $\beta$ increases. Moreover,
$\alpha_{opt}$ is positive (around 0.6), indicating that cooperation
can be optimally enhanced if individuals give large-degree neighbors
more help, but only to a certain extent.

To explain the nonmonotonic behavior displayed in Fig.~\ref{fig2},
we study an individual's payoff as a function of its degree. The
theoretical analysis is provided as follow.

The cost that a cooperator $j$ helps one of its neighbors $i$ can be
calculated as
\begin{equation} \label{eq5}
c_{ji} = \frac{k_{i}^{\alpha}}{\sum_{l\epsilon \Omega_{j}}
k_{l}^{\alpha}} =
\frac{k_{i}^{\alpha}}{k_{j}\sum_{k_{l}=k_{min}}^{k_{max}}P(k_{l}|k_{j})k_{l}^{\alpha}},
\end{equation}
where $P(k_{l}|k_{j})$ is the conditional probability that a node of
degree $k_{j}$ has a neighbor of degree $k_{l}$, $k_{min}$ and
$k_{max}$ are the minimum and maximum node degrees of the network.
Since BA networks have negligible degree-degree
correlation~\cite{dd}, we have approximately
\begin{equation} \label{eq6}
P(k_{l}|k_{j})=k_{l}P(k_{l})/\langle k \rangle,
\end{equation}
where $P(k_{l})$ is the degree distribution of BA networks.
Substituting Eq. (\ref{eq6}) into Eq. (\ref{eq5}), we obtain
\begin{equation} \label{eq7}
c_{ji}=\frac{k_{i}^{\alpha}\langle k \rangle}{k_{j}\langle
k^{\alpha+1} \rangle}.
\end{equation}
According to Eq. (\ref{eq3}) and the mean-field theory, we can write
the payoff $M_{i}$ as
\begin{eqnarray} \label{eq8}
M_{i} &=& -\rho_{c}+r\rho_{c}\sum c_{ji}  \nonumber \\
&=& -\rho_{c}+r\rho_{c}k_{i}\sum_{k_{j}=k_{min}}^{k_{max}}
P(k_{j}|k_{i}) c_{ji}.
\end{eqnarray}
Substituting Eqs. (\ref{eq6}) and (\ref{eq7}) into Eq. (\ref{eq8}),
we obtain
\begin{equation} \label{eq9}
M_{i} = -\rho_{c}+\frac{r\rho_{c}k_{i}^{\alpha+1}}{\langle
k^{\alpha+1} \rangle}.
\end{equation}

From Eq. (\ref{eq9}), one can find that for $\alpha>-1(<-1)$,
higher-degree (lower-degree) individuals gain higher payoffs. In the
case of $\alpha=-1$, individuals with different degree classes have
the same payoffs.

\begin{figure}
\begin{center}
\includegraphics[width=100mm]{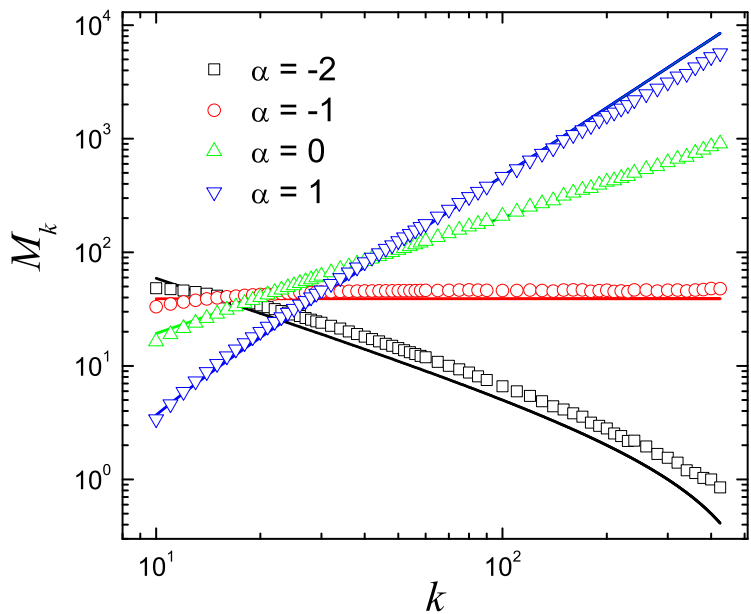}
\caption{(Color online) The payoff $M_{k}$ as a function of degree
$k$ for different values of $\alpha$. The benefit-to-cost ratio $r$
= 40 and the average degree $\langle k \rangle$ = 20. All
individuals are set to be cooperators. The data point are simulation
results and the solid curves are theoretical results from Eq.
(\ref{eq9}).} \label{fig3}
\end{center}
\end{figure}

The quantity $\langle k^{\alpha+1} \rangle$ can be calculated as
$\langle k^{\alpha+1} \rangle
\int_{k_{min}}^{k_{max}}k^{\alpha+1}P(k)dk$, where the degree
distribution of BA networks is $P(k)=2k_{min}^{2}k^{-3}$~\cite{dd}
and the maximum node degree of the network is about
$k_{max}=k_{min}\sqrt{N}$~\cite{db}. After calculating the integral,
we obtain $\langle k^{2} \rangle =k_{min}^{2}\ln{N}$ and $\langle
k^{\alpha+1} \rangle =
2k_{min}^{\alpha+1}(N^{\frac{\alpha-1}{2}}-1)/(\alpha-1)$ (for
$\alpha \neq 1$). From Fig.~\ref{fig3}, one can see that the
theoretical and numerical results are consistent.

For very small values of $\alpha$, lower-degree individuals gain
more payoffs. Besides, the scale-free network is mainly composed of
low-degree nodes. In this case, strategies of low-degree individuals
play important roles in the evolution of cooperation. Note that
cooperators gain less payoff than defectors in the same degree
class. Thus, the whole network will gradually fall into the state of
full defection due to the presence of abundant low-degree defectors.
For large values of $\alpha$, high-degree individuals (so-called
hubs) reap massive profits and gradually become
cooperators~\cite{santos1,santos2}. These hubs and some of their
neighbors will form a cooperator cluster~\cite{Moreno}. Within the
cluster, cooperators can assist each other and the benefits of
mutual cooperation outweigh the losses against the outside
defectors. However, for very large $\alpha$, most individuals inside
the cooperator cluster gain nothing since almost all benefits are
allocated to large-degree individuals. In this case, low-degree
cooperators have negative payoffs since they have to pay the cost of
cooperation. On the contrary, the payoffs of defectors are positive.
As a result, for very large $\alpha$, the cooperator cluster is
vulnerable to the invasion of defectors and become difficult to
expand. Combining the results of the two limits of $\alpha$, the
highest cooperation level should be achieved for some intermediate
values of $\alpha$.

\begin{figure}
\begin{center}
\includegraphics[width=120mm]{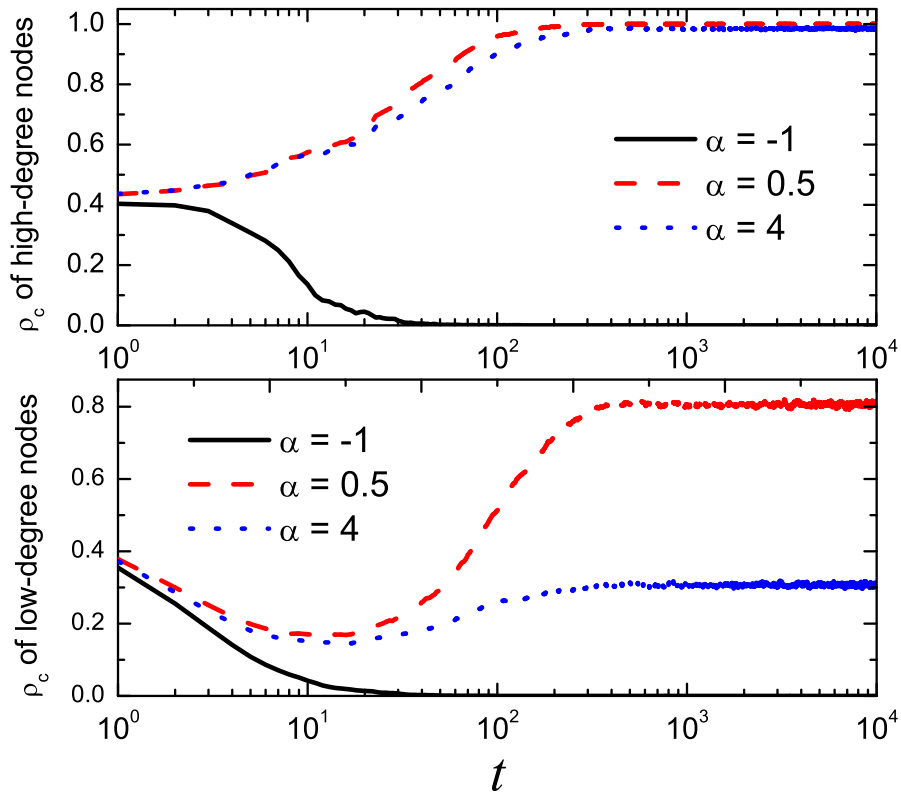}
\caption{(Color online) Time series of the cooperator density
$\rho_{c}$ for high-degree nodes (top panel) and low-degree nodes
(bottom panel) respectively. The average degree $\langle k \rangle$
= 10, the benefit-to-cost ratio $r$ = 2.1 and the noise $\beta$ =
0.5. Without loss of generality, nodes with $k>40$ ($k \leq 40$) are
divided into the high-degree (low-degree) class. For $\alpha=-1$,
both kinds of nodes gradually become defectors. For $\alpha=0.5$,
with time all high-degree nodes become cooperators and most
low-degree nodes choose cooperation. For $\alpha=4$, although all
high-degree nodes finally become cooperators, most low-degree nodes
choose defection in the equilibrium state.} \label{fig4}
\end{center}
\end{figure}

To confirm the above analysis, we divide nodes into two classes:
high-degree and low-degree ones. Then we study the time evolution of
the cooperation density for high-degree and low-degree nodes
respectively. From Fig.~\ref{fig4}, one can see that, for a small
value of $\alpha$ (e.g., $\alpha=-1$), the cooperator density for
both kinds of nodes decreases to 0 as time evolves. For a larger
value of $\alpha$ (e.g., $\alpha=0.5$ or $\alpha=4$), the cooperator
density for high-degree nodes increases to 1 while the cooperator
density for low-degree nodes first decreases and then increases to a
steady value. For $\alpha=0.5$, low-degree nodes inside the
cooperator cluster can gain enough payoffs to resist the invasion of
defectors, leading to a high value of the cooperator density (about
0.8) in the stable state. For $\alpha=4$, low-degree nodes get
little benefit and are vulnerable to the attack of defectors,
resulting in a low cooperation level (about 0.3) in the equilibrium
state.

\begin{figure}
\begin{center}
\includegraphics[width=100mm]{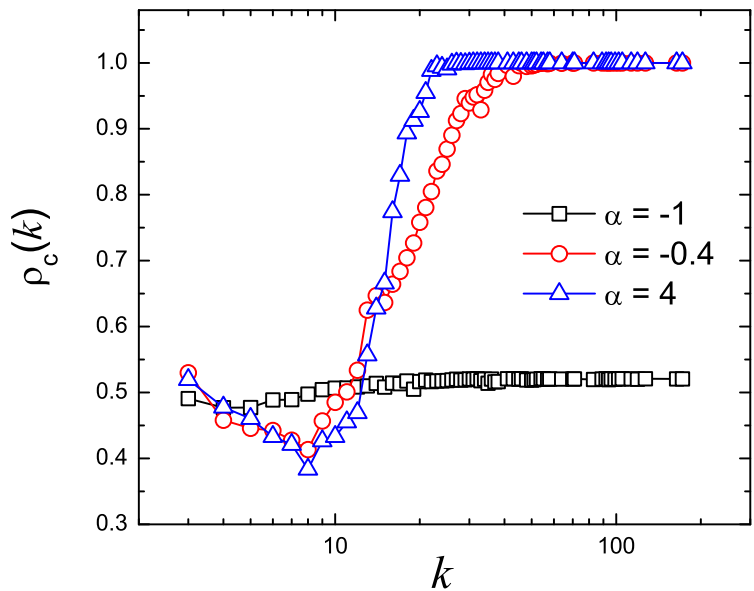}
 \caption{(Color online) The cooperator density $\rho_{c}(k)$ in the steady
state as a function of degree $k$ for different values of $\alpha$.
For each value of $\alpha$, the fraction of cooperators in the
equilibrium state $\rho_{c}$ = 0.5, the average degree $\langle k
\rangle$ = 6 and the noise $\beta$ = 0.5. For $\alpha$ = -1, the
benefit-to-cost ratio $r$ = 3.95. For $\alpha$ = -0.4, $r$ = 1.89.
For $\alpha$ = 4, $r$ = 1.78. For a very small values of $\alpha$
(e.g., $\alpha$ = -1), $\rho_{c}(k)$ is almost independent of degree
$k$. For a large value of $\alpha$ (e.g., $\alpha$ = -0.4 or 4),
high-degree nodes are occupied by cooperators and $\rho_{c}(k)$ is
minimized for medium-degree nodes.} \label{fig5}
\end{center}
\end{figure}

Next, we study the cooperator density $\rho_{c}(k)$ in the steady
state as a function of degree $k$ for different values of $\alpha$.
From Fig.~\ref{fig5}, one can see that for a small value of $\alpha$
(e.g., $\alpha=-1$), $\rho_{c}(k)$ is almost the same for different
values of $k$. For a larger value of $\alpha$ (e.g., $\alpha=-0.4$
or $\alpha=4$), almost all high-degree individuals become
cooperators while some low-degree individuals still choose
defection. Here, we also find that $\rho_{c}(k)$ is minimized for
medium-degree individuals. Such phenomenon has been observed in the
weak prisoner's dilemma game~\cite{min}.

\begin{figure}
\begin{center}
\includegraphics[width=100mm]{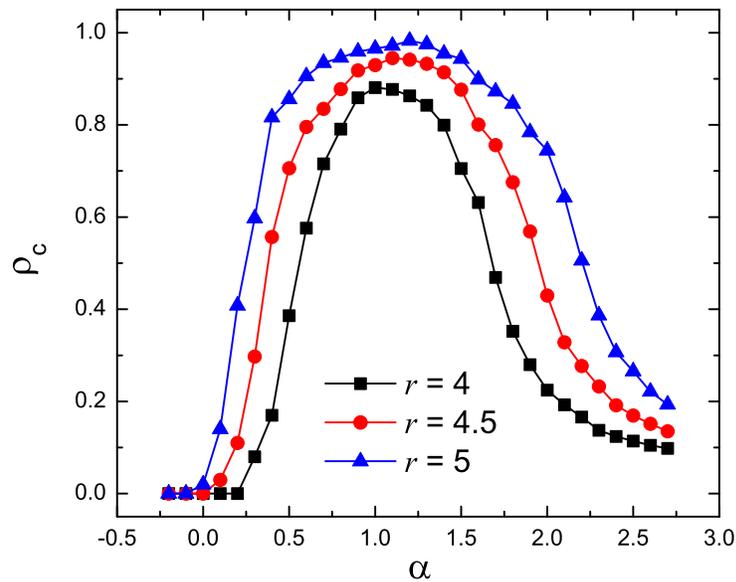}
\caption{(Color online) The fraction of cooperators $\rho_{c}$ as a
function of $\alpha$ for different values of the benefit-to-cost
ratio $r$. The average degree $\langle k \rangle$ = 6 and the noise
$\beta$ = 0.5. The total cost of each cooperator is equal to its
degree, i.e., $c_{i} = k_{i}$. For each $r$, $\rho_{c}$ is maximized
at an optimal value of $\alpha$. } \label{fig6}
\end{center}
\end{figure}

In the above studies, we assume that each cooperator contributes the
same total cost $c_{i}$ = 1. To validate the universality of the
enhancement of cooperation by preferential assistance, we consider a
case in which the total cost of a cooperator is not a constant but
proportional to its degree, i.e., $c_{i} = k_{i}$. In this case,
there also exists an optimal value of $\alpha$, leading to the
highest cooperation level, as shown in Fig.~\ref{fig6}.

\section{Conclusions} \label{sec:discussion}

In conclusion, we have found that cooperation can be promoted when
cooperators contribute more to high-degree neighbors, but only to
some extent. In this case, high-degree individuals are proved to
have high payoffs and act as cooperators. These hubs and some of
their neighbors form a cooperator cluster, within which cooperators
can assist each other and the benefits of mutual cooperation
outweigh the losses against defectors. The above finding is robust
with respect to different values of the benefit-to-cost ratio,
different kinds of network structure, different levels of the noise
to permit irrational choices, and different choices of the total
cost of a cooperator.

The heterogeneous resource allocation has also been considered in
other kinds of game models such as the public goods game. Huang
$et$. $al$. found that cooperation can be enhanced if individuals
invest more to smaller groups~\cite{huang}. Meloni $et$. $al$.
allowed individuals to redistribute their contribution according to
what they earned from the given group in previous
rounds~\cite{Meloni}. Their results showed that not only a Pareto
distribution for the payoffs naturally emerges but also that if
players do not invest enough in one round they can act as defectors
even if they are formally cooperators. Note that the donation game
and the public goods game are based on pair interactions and group
interactions respectively. Together Refs.~\cite{huang,Meloni} and
our work offer a deeper understanding of the impact of the
heterogeneous resource allocation on the evolution of cooperation.

\section*{Acknowledge}
This work was supported by the National Science Foundation of China
(Grant Nos.~61773121, 61403083, 11575072 and 11475074), and the
fundamental research funds for the central universities (Grant No.
lzujbky-2017-172).

\end{document}